\documentclass[reprint,english,superscriptaddress,preprintnumbers,amsmath,amssymb,aps,prx]{revtex4-1}

\usepackage{diagbox}

\usepackage{mathtools}
\usepackage{amsfonts}
\usepackage{mathrsfs}
\usepackage{bbm}
\usepackage{slashed}

\usepackage{graphicx}
\usepackage{color}
\usepackage{soul}
\usepackage{array}

\usepackage{blindtext}
\usepackage{placeins}
\usepackage{booktabs}
\usepackage{makecell}
\usepackage{epstopdf}
\usepackage[caption=false]{subfig}

\usepackage{xspace}
\usepackage{siunitx}
\usepackage{hyperref}
\usepackage[nameinlink]{cleveref}
\usepackage{appendix}

\usepackage{xifthen}
\usepackage{xcolor}
\hypersetup{
	colorlinks,
	linkcolor={red!75!black},
	citecolor={blue!75!black},
	urlcolor={blue!75!black}
}

\setkeys{Gin}{width=0.48\textwidth}

\graphicspath{
	{./figures/}
	{../figures/}
}

\def\s0#1#2{\mbox{\small{$ \frac{#1}{#2} $}}}
\def\0#1#2{\frac{#1}{#2}}

\def\Eq#1{Eq.~\eqref{#1}}

\usepackage{multirow}
\usepackage{colortbl}
\definecolor{kugray5}{RGB}{224,224,224}
\newcommand{\PreserveBackslash}[1]{\let\temp=\\#1\let\\=\temp}
\newcolumntype{C}[1]{>{\PreserveBackslash\centering}p{#1}}
\newcolumntype{R}[1]{>{\PreserveBackslash\raggedleft}p{#1}}
\newcolumntype{L}[1]{>{\PreserveBackslash\raggedright}p{#1}}

\begin{document}
	
	\title{The  effective potential of composite operator  in the first order  region of QCD  phase transition}
	
	\author{Hui-wen Zheng}
	\affiliation{Department of Physics and State Key Laboratory of Nuclear\\ Physics and Technology, Peking University, Beijing 100871, China.}
		
	\author{Yi Lu}
	\affiliation{Department of Physics and State Key Laboratory of Nuclear\\ Physics and Technology, Peking University, Beijing 100871, China.}

	\author{Fei Gao}
	\email{fei.gao@bit.edu.cn}
	\affiliation{School of Physics, Beijing Institute of Technology, 100081 Beijing, China}

\author{Si-xue Qin}
	\email{sqin@cqu.edu.cn}
	\affiliation{Department of Physics and Chongqing Key Laboratory for Strongly Coupled Physics, Chongqing University, Chongqing 401331, China.}
		
	\author{Yu-xin Liu}
	\email{yxliu@pku.edu.cn}
	\affiliation{Department of Physics and State Key Laboratory of Nuclear\\ Physics and Technology, Peking University, Beijing 100871, China.}
	\affiliation{Collaborative Innovation Center of Quantum Matter, Beijing 100871, China.}
	\affiliation{Center for High Energy Physics, Peking University, Beijing 100871, China.}
	
	\date{\today}
	
\begin{abstract}
We propose a method to determine the effective potential of QCD from the gap equation, by introducing the homotopy method between the  solutions of the equation of motion. Via this method, the effective potential beyond the bare vertex approximation is obtained, which then generalizes the Cornwall, Jackiw and Tomboulis (CJT) effective potential for the bilocal composite operators. Moreover, the extended effective potential is set to be a function of self energy instead of the propagator, which is the key point {for the potential to be bounded from below}.
%
%
We then investigate the extended effective potential {in the cases of phase transition 
%
%
of the {QCD} vacuum} with a small current quark mass, and the first-order phase transition of QCD at finite temperature and high baryon chemical potential. 
In the former case, the effective potential shows as an inflection point at the critical mass where the multiple solutions of the Dyson-Schwinger equation (DSE) vanishes, 
which is consistent with that obtained by solving the DSE directly. 
For the latter case, the in-medium properties, such as the latent heat and the difference of trace anomaly, of QCD is obtained. 

	\end{abstract}
	
	\maketitle
	
	\section{\label{sec:level1}Introduction}
	In quantum chromodynamics (QCD), the dynamical chiral symmetry breaking (DCSB) plays a crucial role in the visible mass generation. One of the outcomes of the DCSB is that the masses of low-momentum light quarks experience a significant enhancement of several hundred $\rm{MeV}$ due to the strong interaction. The DCSB can be effectively characterized  by the  composite operators of bilocal quark-antiquark fields. The effective potential for such composite operators has been proposed by Cornwall, Jackiw and Tomboulis (CJT)~\cite{Cornwall:1974vz}. The idea of the CJT effective action is to introduce the bilocal source $J(x,y)$ that couples with the composite operators in the Green function generating functional, and then to take the double Legendre transformation to obtain the CJT effective action, which returns to the conventional effective action when $J(x,y)=0$. The CJT effective action depends not only on the expectation value of the quantum field but also on the expectation value of the composite operators.

However, there are two issues about the CJT effective action. The first is the difficulty of incorporating the equation of motion  beyond the bare vertex approximation to get an explicit expression for the effective action of the respective composite operator.
This is essentially due to the challenge of performing a functional integration with respect to the equation of motion.
The second arises from the fact that the solutions for the CJT effective action are at the saddle point for the composite operators~\cite{Haymaker:1986tt}, which means that the CJT effective action is not bounded from the below,
indicating that  the CJT effective action does not admit a ground state and its vacuum energy is negative infinite.
Now for the second issue, it is caused by the fact that  the  CJT  potential is the function of the bilocal quark-antiquark field, which is proportional to the propagator.  As indicated by the auxiliary field (AF) method~\cite{Haymaker:1986tt,Barducci:1987gn,Casalbuoni:1984du,Casalbuoni:1984ej}, to satisfy the bounded property, the potential should be the function of the self energy instead, \textit{i.e.}, the inverse of the propagator with containing only the sum of all one-particle irreducible diagrams. Therefore, one needs a further transformation which transforms the composite operators in the CJT potential into its inverse counterpart, \textit{i.e.}, the self energy.

We then in this paper propose a method to compute the effective potential from the equation of motion by introducing  the homotopy {transformation}. The effective action is then related to the solutions of the equation of motion, i.e. the Dyson-Schwinger equation (DSE) \cite{Roberts:1994dr,Roberts:2000aa}, of the self-energy, with an additional homotopy parameter.
We then integrate over the homotopy parameter along a certain path in the functional space, to obtain the difference of the potential between the end points of the path. Specifically, the two end points represent two solutions of the DSE.
%
%
We further propose the generating functional for the connected Green function  based on a generalized Legendre transformation,
which converts the variable of CJT effective action, \textit{i.e.}, the quark propagator, to the self energy.
The newly defined effective potential based on such a generalized Legendre transformation
 corresponds to an extension of the AF effective potential, 
 which can naturally return to the AF potential in the case of the bare vertex approximation.

We then study the phase coexistence in QCD, both for the QCD vacuum and for a rather high baryon chemical potential with a first-order phase transition (FOPT).
The QCD phase transitions have been widely studied by the effective models~\cite{Buballa:2003qv,Fukushima:2013rx,Schaefer:2007pw,Schaefer:2008ax,Fu:2007xc,Xin:2014ela},
and the functional QCD method including the DSEs~\cite{Roberts:2000aa,Qin:2010nq,Fischer:2014ata,Fischer:2018sdj,Gunkel:2021oya,Gao:2020qsj,Gao:2020fbl,Gao:2021wun}
, the functional renormalization group (fRG) approach \cite{Fu:2022gou,Fu:2019hdw,Dupuis:2020fhh}, {and the Curci-Ferrari model \cite{Pelaez:2021tpq,Maelger:2019cbk,Reinosa:2015oua}}.
In the QCD vacuum, the gap equation exhibits a chiral-symmetric solution known as the Wigner mode and the DCSB solutions, namely the Nambu$\pm$ modes~\cite{Williams:2006vva,Wang:2012me}.
By utilizing the effective potential, we observe the appearance of a inflection point in the effective potential at a certain current quark mass,
which is caused by the merging of the Nambu$-$ and the Wigner solutions and indicates the disappearance of multiple solutions in its equation of motion.
The critical current quark mass for the merging behaviour in the effective potential {coincides} perfectly with the current quark mass at which the multiple solutions of the quark mass function $M(0)$ vanish.
Additionally, we analyze the QCD phase diagram at finite temperature and density and determine the QCD FOPT line, \textit{i.e.}, on which the pressure between the Nambu$+$ and Wigner phases is equal.
Within the region of the first-order phase transition, we calculate and analyze the latent heat density and the difference of trace anomaly {on the FOPT line}. 

The article is organized as follows: In Section~\ref{sec:2}, the homotopy method for the effective action is presented, and the explicit expression for the effective potential is derived.
In Section~\ref{sec:3}, the quark gap equation is applied to study the effective potential, and a refined truncation scheme for the quark-gluon vertex with the symmetry constraints is presented.
Section~\ref{sec:4} demonstrates the numerical results for the QCD phase diagram, the latent heat density and the difference of trace anomaly and so on.
Finally, Section~\ref{sec:5} summarizes briefly our conclusions and remarks.

\section{Effective potentials of composite operator}
\label{sec:2}

The CJT effective potential approach provides an effective method to characterize the DCSB via the dynamical variables \cite{Cornwall:1974vz},
\textit{i.e.}, the composite operators.
However, it is difficult to apply the CJT effective potential to the DSE beyond the bare vertex approximation.
In this section, a homotopy method is proposed to compute the effective potential from the equation of motion beyond the bare vertex approximation,
where we need at least two solutions of the equation of motion that are related to the homotopy parameter to produce the effective potential.
%
%
We further provide the explicit expression of the generating functional for the inverse form of the equation of motion,
whose Legendre transformation defines the dynamical variables as the self energy, instead of the quark propagator, and therefore its ground state energy is bounded from the below \cite{Haymaker:1986tt}.
In the bare vertex approximation, this newly defined potential is identical to the AF potential, thus it is a natural extension of the AF potential.

	
\subsection{A homotopy method for computing effective potential}
\label{sec:2A}

An effective potential for composite operators has been proposed by Cornwall, Jackiw and Tomboulis (CJT) based on the double Legendre transformation~\cite{Cornwall:1974vz}.
The effective potential depends not only on the field expectation value but also the two point correlation functions, \textit{i.e.}, the propagators.
First, we consider the CJT connected generating functional~\cite{Cornwall:1974vz,Haymaker:1986tt,Barducci:1987gn}:
\begin{align}
	&W_{\rm CJT}\left[ \eta=0 ,\overline{\eta }=0,J=0,K \right]\notag \\& = -\mathrm{Ln} \int{D}(\psi ,{{A}_{\mu }},c)
	\exp \left\{ -\left[ I\left( \psi ,{{A}_{\mu },c}\right) + \overline{\psi }K\psi  \right] \right\}\label{eq:W_CJT}
\end{align}
with $\psi$ the quark field, ${{A}_{\mu }}$ the gauge field, $c$ the ghost field and $K$ bilocal external source, and $I\left( \psi ,{{A}_{\mu }},c \right)$ denotes the classical Euclidean action for the gauge theory.
The shorthand notation is applied as:
\begin{equation}
\overline{\psi }K\psi =\int{{{d}^{4}}x}{{d}^{4}}y{{{\overline{\psi }}}_{\alpha }}(x){{K}_{\alpha \beta }}(x,y){{\psi }_{\beta }}(y)
\end{equation}
where $\alpha$ and $\beta$ collects the color, spin and flavor indices.
	
	
The effective action ${{\Gamma }_{\rm CJT}}[S]$ is the Legendre transform of the generating functional $W_{\rm CJT}[K]$,
which reads~\cite{Cornwall:1974vz}:
\begin{equation}      \label{eq:CJT_action}
{{\Gamma }_{\rm CJT}}[S]=\text{TrLn }\!\![\!\!\text{ }S]-\text{Tr }\!\![\!\!\text{ }SS_{0}^{-1}]-{{\Gamma }_{2}}[S],
\end{equation}
where $S$ is the dressed quark propagator, $S_{0}$ is the free quark propagator, and $\Gamma_{2}[S]$ is the sum of all two-particle-irreducible diagrams expressed in terms of the propagator $S$. The propagator $S$ is then the dynamical variable of the CJT effective action $\Gamma_{\rm CJT}$.
		
The DSE for the quark propagator is related to the first-order functional derivative to the effective action $\Gamma_{\rm CJT}$:
\begin{align}
\frac{\delta {{\Gamma }_{\rm CJT}}\left[ S \right]}{\delta S}=&{{S}^{-1}}-S_{0}^{-1}-\frac{\delta {{\Gamma }_{2}}\left[ S \right]}{\delta S}, \label{eq:cjt_eom}
\end{align}
with the right hand side of Eq.~(\ref{eq:cjt_eom}) being the general form of an equation of motion.
Eq. (\ref{eq:cjt_eom}) also defines the quark self energy  as $\Sigma[S] =\frac{\delta {{\Gamma }_{2}}\left[ S \right]}{\delta S}$, which is the sum over all one-particle-irreducible diagrams for the propagator.

However, the CJT effective action as a function of the propagator in Eq.~(\ref{eq:CJT_action}) only admits saddle points.
To solve this problem, one can apply the Hubbard-Stratonovich transformation to the generating functional Eq.~(\ref{eq:W_CJT}) to obtain the AF generating functional,
whose dynamical variable of the Legendre transformation is the self energy (in the bare vertex approximation), ensuring that the AF effective potential is bounded from below~\cite{Haymaker:1986tt}.
In detail, the AF generating functional is obtained by inserting the functional identity in Eq.~(\ref{eq:W_CJT}):
\begin{align}   \label{eq:au}
\int{\!\!D\left( \Phi  \right)\exp \left[ -\frac{1}{2}\left( \Phi -\psi \bar{\psi } \right)\frac{{{\delta }^{2}}{{\Gamma }_{2}}\left[ S \right]}{\delta {{S}^{2}}}\left( \Phi -\psi \bar{\psi } \right) \right]}=\text{const},
\end{align}
where $\Phi$ is an auxiliary composite field and $\frac{{{\delta }^{2}}{{\Gamma }_{2}}\left[ S \right]}{\delta {{S}^{2}}}$ is the kernel of the four-fermion interaction, which is independent of $S$, indicating that the AF effective potential is in the bare vertex approximation. With some derivation, the AF effective potential can be expressed as \cite{Roberts:1994dr}:	
\begin{equation}   \label{eq:AF_bare}
{{\Gamma }_{{\rm AF}}}[\Sigma]=\text{TrLn }{{[S_{0}^{-1}+\Sigma[S] ]}^{-1}}+\frac{1}{2}\text{Tr}[S\Sigma[S] ],
\end{equation}
where the dynamical variable of the AF action is the self-energy $\Sigma[S]$ and $S=(\frac{{{\delta }^{2}}{{\Gamma }_{2}}\left[ S \right]}{\delta {{S}^{2}}})^{-1} \Sigma$.
In turn, the equation of motion is the first-order functional derivative of ${{\Gamma }_{{\rm AF}}}[\Sigma]$ with respect to $\Sigma$:
\begin{equation}     \label{eq:AF_fo}
\frac{\delta {{\Gamma }_{\rm AF}}\left[ \Sigma  \right]}{\delta \Sigma[S] }=-{{\left( S_{0}^{-1}+\Sigma  \right)}^{-1}}+S.
\end{equation}

Eq.~(\ref{eq:AF_fo}) from the AF potential provides the same quark DSE in Eq.~(\ref{eq:cjt_eom}) as its inverse form,
which therefore leads to a bounded potential.
Now, note that unlike the AF potential itself, the equation of motion in Eq.~(\ref{eq:cjt_eom}) is satisfied even beyond the bare vertex approximation, so we expect Eq.~(\ref{eq:AF_fo}) is also feasible in that case.
Therefore, one may construct the potential upon the equation of motion in Eq.~(\ref{eq:AF_fo}) by taking the self-energy, rather than the propagator, as the dynamical variable of the effective potential.
	
\begin{figure}[htb]
\centering
\includegraphics[width=8.6cm]{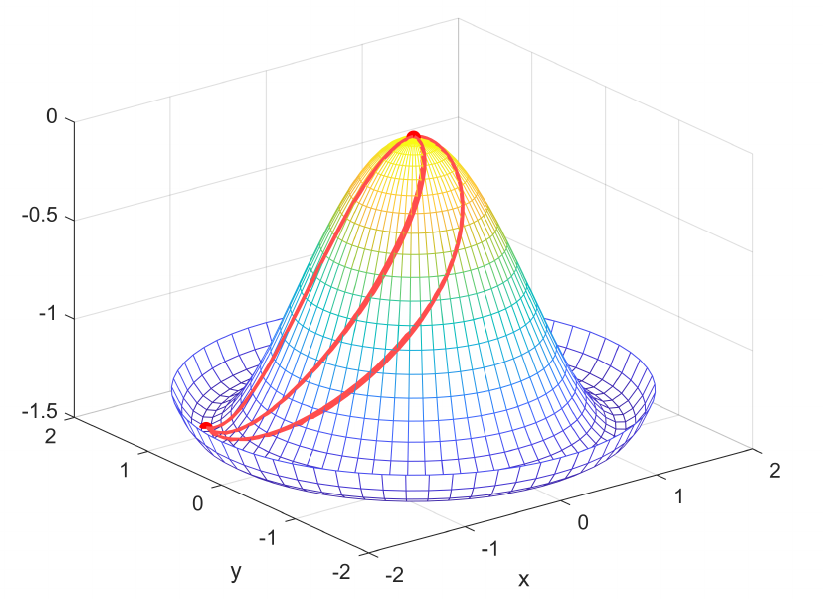}
\caption{A schematic view of homotopy trajectories in the potential.}
\label{mexican_hat}
\end{figure}
		
Here, we propose a homotopy method to determine the value of the effective potential from Eq.~(\ref{eq:AF_fo}) instead of solving the complicated functional integrals directly.
Generally, as shown in Fig.~\ref{mexican_hat}, one can choose different homotopy trajectories in the potential, which shows that the difference between two points in the potential is path independent.
As a pedagogical example, we consider a scalar field $\varphi$ that satisfies the equation of motion $F(\varphi )=0$ where $F(\varphi )$ has an explicit form.
With $F(\varphi )={\delta \Gamma (\varphi )}/{\delta \varphi(x) }$, its effective potential is:
\begin{equation}\label{peda_eq}
\Gamma (\varphi ) = \int{dx\delta \varphi(x) F(\varphi )}.
\end{equation}
Eq.~(\ref{peda_eq}) involves a functional integration which makes it difficult to obtain the value of $\Gamma (\varphi )$ using both analytic and numerical methods directly.
However, one can choose a homotopy trajectory to get the potential $\Gamma (\varphi )$:
\begin{equation}\label{peda_eq1}
\Gamma (\varphi )= \int{d\xi \int{dx}\frac{d\varphi (x)}{d\xi }F(\varphi )},
\end{equation}
where $\xi$ is the homotopy parameter in the chosen trajectory.
Since Eq.~(\ref{peda_eq1}) requires {only} an one-dimensional integral, it is convenient for numerical calculations as it becomes a simplified version of the functional integral.

We will focus on the region where there are two solutions for the DSE, \textit{i.e.}, two extreme points in the potential~\cite{Williams:2006vva,Wang:2012me}.
Considering the homotopy transformation of the gap equation's solutions $\Sigma_{1}$ and $\Sigma_{2}$, we define:
\begin{align}\label{eq:homoS}
\Sigma_{\xi} = \Sigma[S_{\xi}]\,, \quad S_\xi=((1-\xi) S_1^{-1} + \xi S_2^{-1})^{-1},
\end{align}
so that at the end points of $\xi$, the function becomes the propagator solution as:
\begin{align}
\Sigma [S_{\xi=0} = S_{1}] = \Sigma_{1}, \quad \Sigma[S_{\xi=1} = S_{2}] = \Sigma_{2} \,.
\end{align}
If the variation goes along a specific direction, \textit{i.e.}, the direction of the homotopy transformation, it becomes a differential variation.
Then, the difference between any two solutions of the effective potential is given as:
\begin{align}    \label{AF_first_deri}
&{{\Gamma }_{\Sigma}}\left[ {{\Sigma }_{\xi}} \right]-{{\Gamma }_{\Sigma}}\left[ {{\Sigma }_{1}} \right]\notag\\& \; \; = \int_{0}^{\xi} \,d\xi' \text{Tr}\left[ \frac{d\Sigma_{\xi'} }{d\xi' }\frac{\delta {{\Gamma }_{\Sigma}}}{\delta \Sigma_{\xi'} } \right] \notag\\
& \; \;= \int_{0}^{\xi} \,d\xi' \text{Tr}\left\{ \frac{d\Sigma_{\xi'}}{d\xi' } \left[ -{{\left( S_{0}^{-1} + \Sigma_{\xi'} \right)}^{-1}} + {{S}_{\xi' }} \right] \right\}\, .  \;\;
\end{align}
By establishing a specific relation between the parameter $\xi$ and $S$, one can find the difference between any two solutions of the effective action and further determine the effective potential as a function of $\xi$.

\subsection{Extending the CJT effective potential}
\label{sec:2B}

The CJT effective potential has an intrinsic issue, \textit{i.e.}, it has no lower bound. the first attempt to solve the problem was made by applying the Hubbard-Stratonovich transformation to the generating functional.
Consequently, the AF potential with the lower bound was obtained in the bare vertex approximation~\cite{Kleinert:1976xz,Haymaker:1986tt}.
In this work, we construct explicitly an analytical formalism of the effective potential, with its first derivative given in Eq.~(\ref{eq:AF_fo})~\cite{Haymaker:1986tt,Casalbuoni:1984du,Casalbuoni:1984ej,Barducci:1987gn}.
In the bare vertex approximation, our newly established effective potential reduces to the AF effective potential.
This can be achieved in general by a triple Legendre transformation, however, this general triple Legendre transformation to replace the propagator with the self-energy is rather complicated.
The AF method provides an alternative approach, but the previous studies based on the AF method are valid only if the interaction kernel does not depend on the propagator,
which, roughly speaking, is only applicable in the bare vertex approximation and generally contradicts the Slavnov-Taylor identities.
Here we propose a novel form which transforms the propagator dependence of the CJT effective potential into the self-energy dependence.
The form is defined through the generating functional as~\cite{Barducci:1987gn}:

\begin{align}    \label{eq:AF_defi}
{{W}_{\Sigma }}\left[ L \right]=&{{W}_{\rm CJT}}\left[ \eta =0,\overline{\eta }=0,J=0,K \right]\notag\\
&+{{\Gamma }_{2}}\left[ S \right]-{{\Gamma }_{2}}\left[ S+L \right]-\text{Tr}\left[ SK \right].
\end{align}
where $S$ is the propagator, $L$ is the newly introduced external source. The external source of the CJT potential $K$ is then constrained by:
\begin{equation}
K=\Sigma \left[ S \right]-\Sigma \left[ S+L \right].\label{eq:constr}
\end{equation}
Such a condition implies that the external source $K$ is a functional of $L$.
The self-energy is obtained  from the sum of two-particle irreducible potential ${\Gamma }_{2}$, which is defined as $\Sigma \left[ S+L \right]=\frac{\delta {{\Gamma }_{2}}\left[ \left( S+L \right) \right]}{\delta \left( S+L \right)}$.
%

To see how Eq.~(\ref{eq:AF_defi}) transforms the dependence of the dynamical variable $S$ to $\Sigma$, one can recall the Legendre transformation of the CJT generating functional $W_{\rm CJT}[K]$ to ${{\Gamma }_{\rm CJT}}[S]$:
\begin{align}
\frac{\delta {{W}_{\rm CJT}}\left[ K \right]}{\delta K}=& ~S, \label{eq:S_K}\\
{{\Gamma }_{\rm CJT}}\left[ S \right]=& ~{{W}_{\rm CJT}}-\text{Tr}\left[ SK \right],\label{eq:WCJT} \\
\frac{\delta {{\Gamma }_{\rm CJT}}\left[ S \right]}{\delta S}=& -K,\label{eq:CJT_act_defi}
\end{align}
where $S$ is the dressed quark propagator.
This transformation shows that $S$ is the conjugate variable of $K$.
Similarly, the conjugate variable of $L$ in ${{W}_{\Sigma}}\left[ L \right]$ can be found by taking the first-order derivative of ${{W}_{\Sigma}}\left[ L \right]$ in Eq.~(\ref{eq:AF_defi}):
\begin{align}
\frac{\delta {{W}_{\Sigma}}\left[ L \right]}{\delta L}=&\ S\frac{\delta K}{\delta L}+\Sigma \left[ S \right]\frac{\delta S}{\delta L}-\Sigma \left[ S+L \right]\frac{\delta S}{\delta L} \notag \\
& -\Sigma \left[ S+L \right]-\frac{\delta S}{\delta L}K-S\frac{\delta K}{\delta L}, \notag\\
= & \left( \Sigma \left[ S \right]-\Sigma \left[ S+L\right]-K \right)\frac{\delta S}{\delta L}-\Sigma \left[ S+L \right],\label{eq:W_K}
\end{align}
where Eq.~(\ref{eq:S_K}) has been utilized in the first line.
Using the constraint of $K$ in \Eq{eq:constr}, we can eliminate the {first term} in Eq. (\ref{eq:W_K}) with only $-\Sigma\left[ S+L \right]$ left,
which means that the dynamical variable of the effective potential ${{\Gamma}_{\Sigma}}$ corresponding to ${{W}_{\Sigma}}$ is the self-energy $\Sigma[S+L]$.
Therefore, only the constraint condition in Eq.~(\ref{eq:constr}) can ensure that the self-energy is the dynamical variable,
otherwise, the first term in Eq.~(\ref{eq:W_K}) cannot be eliminated and the dynamical variables of the effective potential would be a mixture of the self-energy and the propagator,
making it difficult for further derivations.
In fact, it is precisely the absence of Eq.~(\ref{eq:constr}) that prevents the Legendre transformation of the AF connected generating functional from deriving a potential with the self-energy as the variable beyond the bare vertex case.
Therefore, we can define the Legendre transform of $W_{\Sigma}$ as:
\begin{align}
\frac{\delta {W_{\Sigma}}\left[ L \right]}{\delta L}=&-\Sigma \left[ S+L \right], \\
{{\Gamma }_{\Sigma}}\left[ \Sigma  \right]=&~ {W_{\Sigma}}+\text{Tr}\left[ \Sigma L \right],\label{eq:AF_eff} \\
\frac{\delta {{\Gamma }_{\Sigma}}\left[ \Sigma  \right]}{\delta \Sigma }=& ~L,\label{eq:AF_moe}
\end{align}
where Eq.~(\ref{eq:AF_eff}) identifies the  dynamical variable in $\Gamma_{\Sigma}$ as $\Sigma \left[ S+L \right]$.
We can utilize the CJT series $\Gamma_{\rm CJT}$ in Eq.~(\ref{eq:CJT_action}) and $W_{\rm CJT}$ in Eq.~(\ref{eq:WCJT}) to rewrite the newly defined connected generating functional $W_{\Sigma}$ in Eq.~(\ref{eq:AF_defi}) as:
\begin{align}
{{W}_{\Sigma }}\left[ L \right]=& \text{TrLn }[S]-\text{Tr }[SS_{0}^{-1}]-{{\Gamma }_{2}}[S]+\text{Tr}\left[ S K \right]\notag \\
& +{{\Gamma }_{2}}\left[ S \right]-{{\Gamma }_{2}}\left[ S+L \right]-\text{Tr}\left[ S K \right]\notag \\
=& \text{TrLn }[S]-\text{Tr }[S S_{0}^{-1}]-{{\Gamma }_{2}}\left[ S+L \right] ,\label{eq:WAF}
\end{align}
where one can find the external source $L$ appearing in the sum of two-particle irreducible graphs $\Gamma_2$. With Eq.~(\ref{eq:WAF}) and Eq.~(\ref{eq:AF_eff}) we get:
\begin{align}
{{\Gamma }_{\Sigma}}\left[ \Sigma  \right]=& \text{TrLn}\left[ S \right]-\text{Tr}\left[ S S_{0}^{-1} \right]  -{{\Gamma }_{2}}\left[ {{ S+L }} \right]+\text{Tr}\left[ \Sigma L \right]. \label{Gamma_AF}
\end{align}
	
By taking the first functional derivative of $\Gamma_{\Sigma}$ with respect to $\Sigma$ in Eq.~(\ref{eq:AF_moe}):
\begin{align}    \label{deri_Gamma}
\frac{\delta {{\Gamma }_{\Sigma }}}{\delta \Sigma }= & \, {{S}^{-1}}\frac{\delta S}{\delta \Sigma }-\frac{\delta S}{\delta \Sigma }S_{0}^{-1}\notag \\
& \, -\frac{\delta \ {{\Gamma }_{2}}[S+L]}{\delta (S+L)}\frac{\delta (S+L)}{\delta \Sigma }+L+\Sigma \frac{\delta L}{\delta \Sigma } \notag\\
= & \, {{S}^{-1}}\frac{\delta S}{\delta \Sigma }-\frac{\delta S}{\delta \Sigma }S_{0}^{-1}-\frac{\delta \ {{\Gamma }_{2}}[S+L]}{\delta (S+L)}\frac{\delta S}{\delta \Sigma }+L \notag\\
= & \, ({{S}^{-1}}-S_{0}^{-1}-\Sigma [S+L])\frac{\delta S}{\delta \Sigma }+L \, .
\end{align}
The equation of motion of $\Gamma_{\Sigma}$ with external source $L$ is then obtained by comparing Eq. (\ref{deri_Gamma}) to Eq.~(\ref{eq:AF_moe}) as:
\begin{equation}     \label{eq:eom_Gamma}
{{S}^{-1}}-S_{0}^{-1} - \Sigma [S+L] = 0 \, .
\end{equation}

Note that Eq.~(\ref{eq:eom_Gamma}) is the generalized equation of motion with nonvanishing external source $L$. By inserting $K=\Sigma \left[ S \right]-\Sigma \left[ S+L \right]$ into Eq.~(\ref{eq:eom_Gamma}), we have:
\begin{equation}   \label{eq:eom}
{{S}^{-1}}-S_{0}^{-1}-\Sigma [S]=-K,
\end{equation}
Eq.~(\ref{eq:eom}) returns to the form of the general CJT equation of motion, which comes from comparing Eq.~(\ref{eq:cjt_eom}) and Eq.~(\ref{eq:CJT_act_defi}).
	
With the generalized equation of motion in Eq.~(\ref{eq:eom_Gamma}), one can express the $\Gamma_{\Sigma}$ in Eq.~(\ref{Gamma_AF}) as:
\begin{align}    \label{eq:AF_sigma}
{{\Gamma }_{\Sigma}}\left[ \Sigma  \right]=&~\text{TrLn}{{\left[ S_{0}^{-1}+\Sigma  \right]}^{-1}}-\text{Tr}\left[ 1 \right]\notag\\&-{{\Gamma }_{2}}\left[ {{\left( S+L \right)}_{ }} \right]
+\text{Tr}\left[ {{\left( S+L \right)}_{ }}\Sigma  \right] .
\end{align}
By taking the first functional derivative of Eq.~(\ref{eq:AF_sigma}) with respect to $\Sigma$, we find:
\begin{align}   \label{eq:deri_AF}
\frac{\delta {{\Gamma }_{\Sigma}}\left[ \Sigma  \right]}{\delta \Sigma }=& \, -{{\left( S_{0}^{-1}+\Sigma  \right)}^{-1}}-\frac{\delta \ {{\Gamma }_{2}}[S+L]}{\delta (S+L)}\frac{\delta (S+L)}{\delta \Sigma } \notag\\
& \; +(S+L)+\Sigma \frac{\delta (S+L)}{\delta \Sigma } \notag\\
=& \, -{{\left( S_{0}^{-1}+\Sigma \right)}^{-1}}+\left( S+L \right),
\end{align}
where $\Sigma=\Sigma[S+L]$. Eq.~(\ref{eq:AF_fo}) is then obtained by replacing $S+L$ with $S$.
	

Finally, we show that the newly defined effective potential $\Gamma_{\Sigma }$ in Eq.~(\ref{eq:AF_sigma}) is the natural extension of the AF effective potential.
With the bare vertex approximation, \textit{i.e.}, $\Gamma_{2}$ being quadratic to $(S+L)$, this potential returns to the AF effective potential~\cite{Haymaker:1986tt,Barducci:1987gn}:
\begin{align}
{{\Gamma }_{\Sigma}}\left[ \Sigma  \right]= & ~\text{TrLn}{{\left[ S_{0}^{-1}+\frac{{{\delta }^{2}}{{\Gamma }_{2}}}{\delta {{(S+L)}^{2}}}\left( S+L \right) \right]}^{-1}}-\text{Tr}\left[ 1 \right]\notag \\
& \; +\frac{1}{2}\text{Tr}\left[ \left( S+L \right)\frac{{{\delta }^{2}}{{\Gamma }_{2}}}{\delta {{(S+L)}^{2}}}\left( S+L \right) \right],
\end{align}
which is identical to Eq.~(\ref{eq:AF_bare}) by replacing $S+L$ with $S$.
	
The relation between the AF and the CJT connected generating functional {was} originally obtained  by applying the same Hubbard-Stratonovich transformation in Eq.~(\ref{eq:au}) to the CJT generating function in \Eq{eq:W_CJT}
and then integrating the fermion fields out:
\begin{widetext}
\begin{equation}
{{W}_{\rm CJT}}\left[ K \right]=-\mathrm{Ln}\mathop{\int }^{}D\left( \Phi  \right)\exp \left\{ -\left[ -\text{Trln}\left( S_{0}^{-1}+\frac{{{\delta }^{2}}{{\Gamma }_{2}}\left[ S \right]}{\delta {{S}^{2}}}\Phi +K \right) +\frac{1}{2}\Phi\frac{{{\delta }^{2}}{{\Gamma }_{2}}\left[ S \right]}{\delta {{S}^{2}}}\Phi  \right] \right\},\label{eq:CJT_au}
\end{equation}
\end{widetext}
%
	
After changing the integrating variable $\Phi \to \Phi -{{\left( \frac{{{\delta }^{2}}{{\Gamma }_{2}}\left[ S \right]}{\delta {{S}^{2}}} \right)}^{-1}}K$ in Eq.~(\ref{eq:CJT_au}),
one can find the relation between the CJT and the AF connecting generating functional as~\cite{Haymaker:1986tt,Barducci:1987gn,Casalbuoni:1984du,Casalbuoni:1984ej}:
\begin{equation}   \label{eq:AF_2}
{{W}_{\rm CJT}}={{W}_{\rm AF}}+\frac{1}{2}K{{\left( \frac{{{\delta }^{2}}{{\Gamma }_{2}}\left[ S \right]}{\delta {{S}^{2}}} \right)}^{-1}}K.
\end{equation}
In case of the bare vertex approximation, one can also see the newly defined generating functional, \textit{i.e.} Eq.~(\ref{eq:AF_defi}), obeys the same relation Eq.~(\ref{eq:AF_2}) as the AF generating functional. The bare vertex approximation suggests that the $\Gamma_{2}[S]$ is quadratic of $S$:
\begin{equation}\label{Gamma2_bare}
    {{\Gamma }_{2}}[S]=\frac{1}{2}S\frac{{{\delta }^{2}}{{\Gamma }_{2}}[S]}{\delta {{S}^{2}}}S=\frac{1}{2}S\frac{\delta \, {{\Gamma }_{2}}[S]}{\delta S},
\end{equation}
Also, Eq.~(\ref{eq:constr}) shows a constraint as $K=-\frac{{{\delta }^{2}}{{\Gamma }_{2}}}{\delta {{S}^{2}}}L$, which yields:
\begin{align}\label{SK}
  \text{Tr}\left[ SK \right]& =-SL\frac{{{\delta }^{2}}{{\Gamma }_{2}}\left[ S \right]}{\delta {{S}^{2}}} \notag\\ 
 & =-L\frac{\delta {{\Gamma }_{2}}\left[ S \right]}{\delta S},  
\end{align}
where the second line applies Eq.~(\ref{Gamma2_bare}).  Then, 
One can expand Eq.~(\ref{eq:AF_defi}) up to second order of $L$ and apply the constraint $K=-\frac{{{\delta }^{2}}{{\Gamma }_{2}}}{\delta {{S}^{2}}}L$ and Eq.~(\ref{SK}):
\begin{align}\label{eq:new_CJT}
  {{W}_{\Sigma }}\left[ L \right]& ={{W}_{\text{CJT}}}\left[ K \right]+{{\Gamma }_{2}}\left[ S \right]-{{\Gamma }_{2}}\left[ S+L \right]-\text{Tr}\left[ SK \right] \notag\\ 
 & ={{W}_{\text{CJT}}}\left[ K \right]-\frac{1}{2}L\frac{{{\delta }^{2}}{{\Gamma }_{2}}\left[ S \right]}{\delta {{S}^{2}}}L \notag\\ 
 & ={{W}_{\text{CJT}}}\left[ K \right]-\frac{1}{2}K{{\left( \frac{{{\delta }^{2}}{{\Gamma }_{2}}\left[ S \right]}{\delta {{S}^{2}}} \right)}^{-1}}K.  
\end{align}
The connected generating functional is then simplified to Eq.~(\ref{eq:AF_2}).
The above derivation demonstrates that our newly defined connected generation functional in Eq.~(\ref{eq:AF_defi}) is a natural extension of the AF effective potential. 

 \begin{figure}[htbp]
\centering
\includegraphics[width=7.5cm]{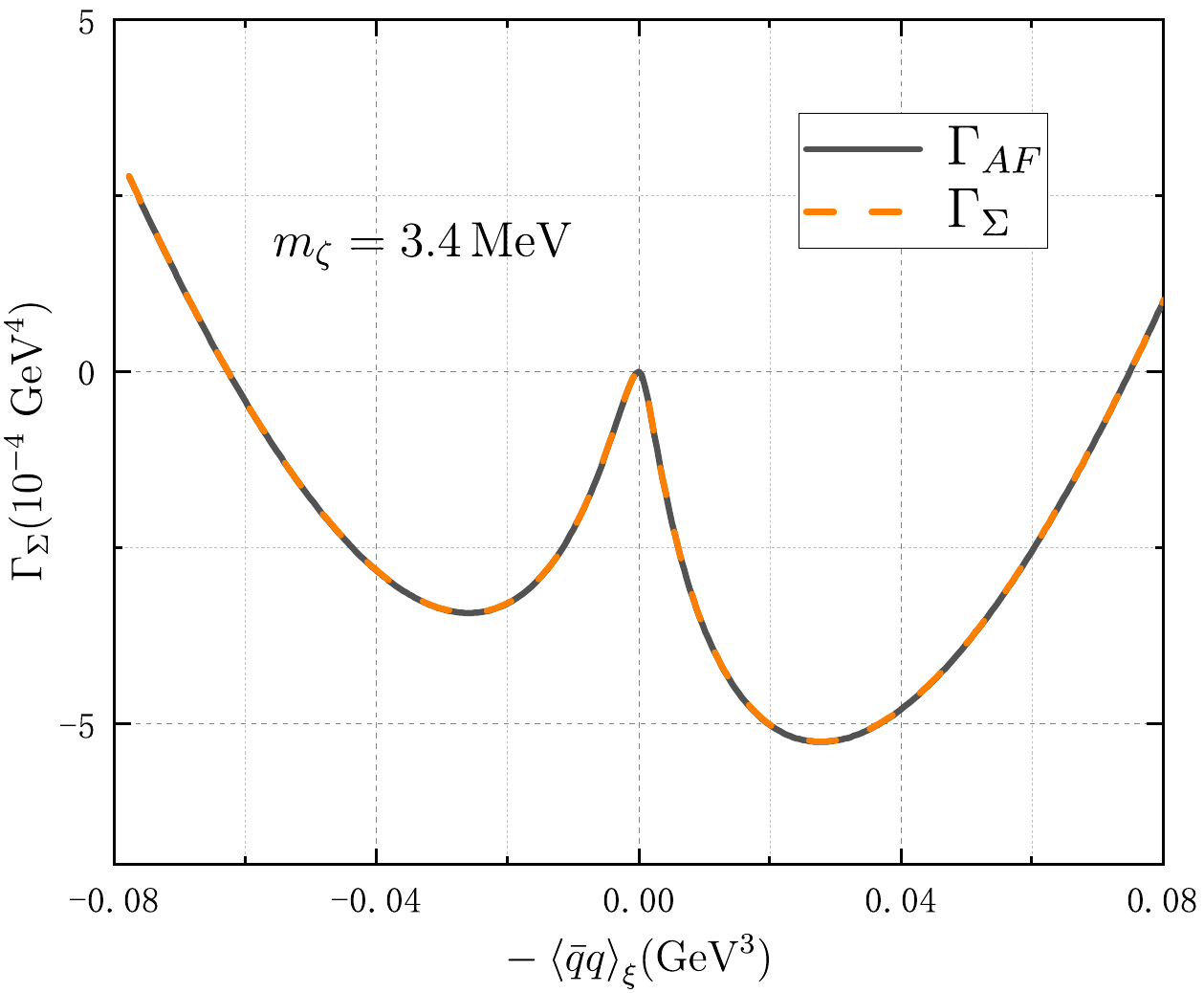}
\vspace*{-3mm}
\caption{The comparision between the AF potential and the newly defined potential, both of which are in the bare vertex aprroximation.}
\label{fig:eff_compare}
\end{figure}

Moreover, Eq.~(\ref{eq:AF_2}) implies an incomplete transformation from the CJT to the AF connecting generation functional, up to the quadratic term of the source, which prevents us from obtaining the explicit expression of the equation of motion of the AF connecting generating functional, making it difficult to employ the homotopy method to the AF potential. We then compare numerically the newly defined potential with the AF potential in the case of bare vertex approximation in Fig.~\ref{fig:eff_compare}, where the AF potential is obtained from Eq.~(\ref{eq:AF_bare}), and $\Gamma_\Sigma$ is obtained from the homotopy method, \textit{i.e.} Eq.~(\ref{AF_first_deri}).
Then, we find that there is no difference between two potentials in Fig.~\ref{fig:eff_compare}, which is consistent with the math proof in the Eq.~(\ref{eq:new_CJT}) and Eq.~(\ref{eq:AF_2}).

 In conclusion, the newly defined effective potential is  a generalized Legendre transform of the CJT effective potential and is consistent with the DSE in  any kind of truncation scheme. Therefore, with the homotopy method, one can now directly compute the effective potential starting from a given DSE with a specific choice of \textit{Ans\"atze}.

\section{The scheme of Dyson-Schwinger equations}
\label{sec:3}



To study the QCD phase transition at finite temperature and density with the potential, one needs to obtain the dressed quark propagator at finite temperature and chemical potential,
whose behavior is controlled by the  quark gap equation, which reads:
\begin{align}    \label{eq:gap}
{{S}^{-1}}({{{\tilde{\omega }}}_{n}},\vec{p})={{Z}_{2}}(i{{\gamma }_{4}}{{{\tilde{\omega }}}_{n}}+i\vec{\gamma }\cdot \vec{p})+{{Z}_{4}}{{m}_{\zeta }}+\Sigma ({{{\tilde{\omega }}}_{n}},\vec{p})
\end{align}
with self energy:
\begin{align}
\Sigma ({{{\tilde{\omega }}}_{n}},\vec{\,p})=& \, \frac{4}{3}{{g}^{2}}{{Z}_{1}}T\sum\limits_{m=-\infty }^{\infty }{\int_{q}^{\Lambda }{{{D}_{\mu \nu }}}({{\Omega }_{nm}},\vec{k};T,{{\mu }_{q}})} \notag\\
& \, \times {{\gamma }_{\mu }}S({{{\tilde{\omega }}}_{m}},\vec{q}){{\Gamma }_{\nu }}({{{\tilde{\omega }}}_{n}},\vec{\,p},{{{\tilde{\omega }}}_{m}},\vec{\,q};T,{{\mu }_{q}} ) \, ,
\end{align}
which is defined in the Euclidean space~\cite{Roberts:1994dr},
with $\mu_{q}$ the quark chemical potential,
${{\omega }_{n}}=(2n+1)\pi T$ the Matsubara frequency and ${{{\tilde{\omega }}}_{n}}={{\omega }_{n}}+i{{\mu }_{q}}$ for the quarks.
${{\Omega }_{nm}}={{\omega }_{n}}-{{\omega }_{m}}$ is the Matsubara frequency of the gluon.
$Z_1,\,Z_2,\, Z_m$ are renormalisation constants of the quark-gluon-vertex, quark wave function and quark mass, respectively, with also the definition for $Z_{4} = Z_{2} Z_{m}$.
$m_{\zeta}$ is the current quark mass at the renormalization point $\zeta$,
$\int_{\vec{q}}^{\Lambda}\coloneqq \int _0^{\Lambda} d^3q/(2\pi)^3$ with $\Lambda$ the energy scale of the loop momentum regularisation.
Finally, $\Gamma_{\nu}$ is the dressed quark-gluon interaction vertex and $D_{\mu\nu}^{ab}$ is the dressed gluon propagator. The general form of the solution of the gap equation can be expressed as:
\begin{align}
{{S}^{-1}}({{{\tilde{\omega }}}_{n}},\vec{p})=& \, i\vec{\gamma }\cdot \vec{p}A({{{\tilde{\omega }}}_{n}},\vec{p}) \notag\\
& \, +i{{\gamma }_{4}}{{{\tilde{\omega }}}_{n}}C({{{\tilde{\omega }}}_{n}},\vec{p})+B({{{\tilde{\omega }}}_{n}},\vec{p}).
\end{align}
	

The DSEs of QCD provide an infinite tower of the coupled equations, so a truncation is required to make the gap equation a closed system.
Here we utilize the functional-lattice fit of the gluon propagator and the  \textit{Ans\"atze} of the quark-gluon interaction vertex in Eq.~(\ref{eq:gap}).
To represent the O(4) symmetry breaking in case of finite temperature and chemical potential, the dressed gluon propagator takes the form:
\begin{align}
{{D}_{\mu \nu }}({{\Omega }_{nm}},\vec{k})=P_{\mu \nu }^{T}{{D}_{T}}({{\Omega }_{nm}},\vec{k})+P_{\mu \nu }^{L}{{D}_{L}}({{\Omega }_{nm}},\vec{k}),
\end{align}
where $P_{\mu \nu }^{L,T}$ are the longitudinal and transverse projection operators:
\begin{equation}\label{projct}
\begin{split}
& P_{\mu \nu }^{T}=(1-{{\delta }_{\mu 4}})(1-{{\delta }_{\nu 4}})({{\delta }_{\mu \nu }}-\frac{{{k}_{\mu }}{{k}_{\nu }}}{{{k}^{2}}}), \\
& P_{\mu \nu }^{L}=({{\delta }_{\mu \nu }}-\frac{{{k}_{\mu }}{{k}_{\nu }}}{{{k}^{2}}})-P_{\mu \nu }^{T},
\end{split}
\end{equation}
where $k=({{\Omega }_{nm}},\vec{k})$.
	
In the case of finite temperature and density, the gluon acquires a screening mass from the quark loop and the gauge loops, the latter of which can be approximated by the temperature dependence of the quenched gluon propagator $D_{L/T,que}$,
one has then:
\begin{align}
& D_{L}^{-1}=D_{vac}^{-1}+m_{q}^{2}+D_{L,que}^{-1}(T)-D_{L,que}^{-1}(T=0), \\
& D_{T}^{-1}=D_{vac}^{-1}+D_{T,que}^{-1}(T)-D_{T,que}^{-1}(T=0),
\end{align}
with the screening mass from quarks:
\begin{equation}
m_{q}^{2}=4\pi {{\alpha }_{q\bar{q}A}} \, N_{f}^{g} \left( \frac{{{T}^{2}}}{6}+\frac{\mu _{q}^{2}}{2{{\pi }^{2}}} \right), \label{eq:htl-qrk}
\end{equation}
where $N_{f}^{g}=3$ is the active flavour number, and the temperature dependence of quenched gluon propagator can be directly read off from the lattice QCD computation~\cite{Maas:2011ez,Eichmann:2015kfa}.
We also apply the functional-lattice fit of the dressed gluon propagator for 2+1 flavor in the vacuum, which reads~\cite{Gao:2021wun,Aguilar:2019uob,Boucaud:2018xup,Gao:2020qsj,Gao:2020fbl}:
\begin{align}   \label{lattice gluon}
D_{vac}(p)=\frac{(a^2+p^2)/(b^2+p^2)}{M_G^2(p^2)+p^2[1+c\,\mathrm{ln}(d^2p^2+e^2M^2(p^2))]^\gamma},
\end{align}
with
\begin{align}    \label{eq:M2}
M_G^2(p^2)=f^4/(g^2+p^2),
\end{align}
where $\gamma=(13-4/3\,N_f)/(22-4/3\,N_f)$ is the anomalous dimension for one-loop gluon propagator, and $N_f=4$ is chosen here. The fit parameters are $[a, b, c, d, e]=[1\,\mathrm{GeV}, 0.735 \,\mathrm{GeV}, 0.12, 0.0257\, \mathrm{GeV}^{-1}, 0.081\, \mathrm{GeV}^{-1}]$, and $[f, g]=[0.65\, \mathrm{GeV}, 0.87\, \mathrm{GeV}]$. This gluon propagator is valid for a large regime for $p\in [0,40] \mathrm{GeV}$. We then set the renormalization point at $\zeta=12\,\mathrm{GeV}$ and apply the multiplicative renormalization for the quark and  gluon propagator, and also the ghost dressing in the quark gluon vertex. The coupling constant $\alpha_{q\bar{q}A}=0.2255$ is then adopted here which is consistent with the previous full computation.
	
Generally speaking, by taking the first-order derivative of the new effective potential in Eq.~(\ref{eq:AF_sigma}), we can derive the DSE without any approximation. Therefore, to make the quark DSE closed, we must impose additional constraints from other aspects, such as QCD gauge symmetries. We then take the minimal \textit{Ans\"atze} of the dressed quark-gluon interaction vertex as~\cite{Gao:2020qsj}:
\begin{equation}          \label{vertex: qg}
\begin{split}
{{\Gamma }_{\mu }}({{{\tilde{\omega }}}_{n}},\vec{p},{{{\tilde{\omega }}}_{m}},\vec{q})=& \, F({{k}^{2}})\frac{A({{{\tilde{\omega }}}_{n}},\vec{p})+A({{{\tilde{\omega }}}_{m}},\vec{q})}{2}{{\gamma }_{\mu }} \\
& +Z_{A,L}^{-1/2}({{k}^{2}}){{\Delta }_{B}}P_{\mu \nu }^{L}{{\sigma }_{\mu \nu }}{{k}_{\nu }} \\
& +Z_{A,T}^{-1/2}({{k}^{2}}){{\Delta }_{B}}P_{\mu \nu }^{T}{{\sigma }_{\mu \nu }}{{k}_{\nu }}
\end{split}
\end{equation}
with
%
\begin{align}
& {{\Delta }_{B}}=\frac{\tilde{B}({{{\tilde{\omega }}}_{n}},\vec{p})-\tilde{B}({{{\tilde{\omega }}}_{m}},\vec{q})}{{{{\tilde{\omega }}}_{n}}^{2}+{{{\vec{p}}}^{2}}-{{{\tilde{\omega }}}_{m}}^{2}-{{{\vec{q}}}^{2}}}, \\
& \tilde{B}({{{\tilde{\omega }}}_{n}},\vec{p})=B({{\omega }_{0}}~\text{sgn}({{\omega }_{\text{n}}})+i{{\mu }_{q}},{{l}_{p}}),
\end{align}
{where $k=({{\Omega }_{nm}},\vec{q}-\vec{p})$, ${{l}_{p}}={{({{{\vec{p}}}^{2}}+{{\omega }_{n}}^{2}-{{\omega }_{0}}^{2})}^{1/2}}$}, ${{Z}_{A,L}}={{D}_{L}}({{k}^{2}}){{k}^{2}}$ and ${{Z}_{A,T}}={{D}_{T}}({{k}^{2}}){{k}^{2}}$ are the gluon dressing functions , and $F(k^2)$ is the ghost dressing function.
The first term in Eq.~(\ref{vertex: qg}) is a generalized   Ball-Chiu \textit{Ans\"atze}~\cite{Ball:1980ay,Qin:2016fbu} which is based on the Slavnov-Taylor Identities (STI) with only the 2-point correlation functions considered.
The remaining terms are the next dominant terms to the dynamical mass since they are related to the anomalous magnetic moment of quark and its relation with the quark dressing function $B$ has been observed from the transverse Ward-Green-Takahashi identity (tWTGI)~\cite{Chang:2010hb,Qin:2013mta,Gao:2020qsj}.
In short, the quark-gluon vertex is constrained from the QCD gauge symmetries, which self-consistently incorporates the 2PI contribution in the new potential.  
 We then extract the numerial results for the corresponding potential by applying the homotopy method to DSE.
	
We also give an analytical form for the ghost dressing function, which is fitted from the functional renormalization group (FRG) data as~\cite{Cyrol:2017ewj}:
\begin{align}
F({{k}^{2}})=\frac{\left( {{a}_{1}}+{{b}_{1}}\sqrt{{{k}^{2}}}+{{k}^{2}} \right)/\left( {{c}_{1}}+{{d}_{1}}\sqrt{{{k}^{2}}}+{{k}^{2}} \right)}{{{\left[ 1+{{e}_{1}}\ln \left( f_{1}^{2}{{k}^{2}}+g_{1}^{2}\ {{M}_G^{2}}\left( {{k}^{2}} \right) \right) \right]}^{\delta }}},
\end{align}
where $M_G^2(p^2)$ is defined in Eq.~(\ref{eq:M2}).
The fitted parameters are $\left[ {{a}_{1}},\ {{b}_{1}},\ {{c}_{1}},\ {{d}_{1}},\ {{e}_{1}},\ {{f}_{1}},\ {{g}_{1}} \right]=[0.152\ \text{Ge}{{\text{V}}^{\text{2}}},\ 0.697\ \text{GeV},\ 0.0055\ \text{Ge}{{\text{V}}^{\text{2}}},0.016\ \text{GeV},0.045$ $, (0.025\ \text{GeV})^{-2}, (0.0237\ \text{GeV})^{-2}]$.
In the ultraviolet region, the anomalous dimension of the one-loop ghost dressing function is constrained by the gluon dressing function as~\cite{Fischer:2002hna,Fischer:2003rp} $\delta  = \left( {1 - \gamma } \right)/2=0.27$.
		
\begin{figure}[htb]
\centering
\includegraphics[width=7.5cm]{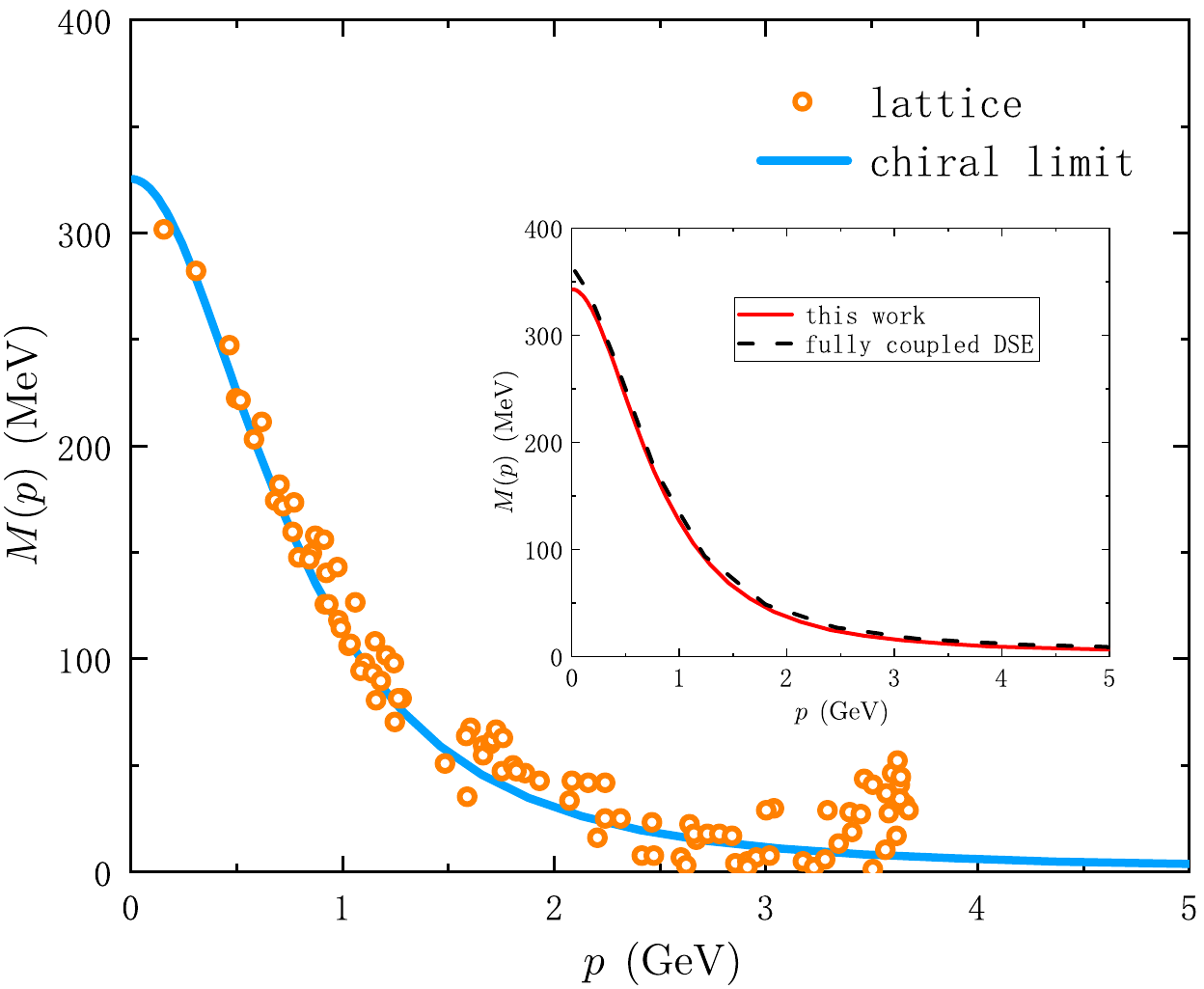}
\vspace*{-4mm}
\caption{The quark mass function in the chiral limit calculated with the minimal \textit{Ans\"atze} and the comparison with the lattice QCD results in vacuum: Blue-solid - the minimal \textit{Ans\"atze} solution, orange-dot - the lattice QCD results from Ref. \cite{Bowman:2005vx} extrapolated to the chiral limit case. The inset shows the comparison between the mass functions obtained from the minimal \textit{Ans\"atze} and from the fully-coupled DSE \cite{Gao:2021wun}, at the physical current quark mass.}
		\label{mass_fun_lattice}
\end{figure}

With this truncation scheme, we obtain the quark mass function in the vacuum,
whose running behavior is well comparable with the lattice QCD simulation results and those from the full computation of the functional QCD approaches~\cite{Bowman:2005vx,Gao:2021wun},
as depicted  in Fig.~\ref{mass_fun_lattice}. The order parameter of the DCSB, \textit{i.e.}, the quark condensate, can then be obtained as:
\begin{align}    \label{q_Cd}
- {\left\langle {\bar qq} \right\rangle _{m,\zeta }} =& \, {N_c}{Z_4}({\zeta ^2},{\mkern 1mu} {\Lambda ^2}) \notag\\
&\times {{\rm{tr}}_{\rm{D}}}\int_{0}^{\Lambda}  {\frac{{{d^4}q}}{{{{\left( {2\pi } \right)}^4}}}\frac{{B\left( {{q^2}} \right)}}{{{q^2}{A^2}\left( {{q^2}} \right) + {B^2}\left( {{q^2}} \right)}}},
\end{align}
where $\mathrm{tr_{D}}$ identifies only the trace over the Dirac indices. For an asymptotically enough renormalization scale $\zeta$,
the quark condensate is related to renormalization-group-invariant (RGI) quark condensate with~\cite{Roberts:1994dr,Chen:2021ikl}:
\begin{align}          \label{RGI}
\left \langle \bar{q}q \right \rangle_{m,\zeta} = \left \langle \bar{q}q \right \rangle_{m} \big{(} \frac{1}{2}\mathrm{ln}\,\zeta^2/\Lambda_{\rm QCD}^{2} \big{)}^{\gamma_{m}} ,
\end{align}
with $\gamma_{m} = 12/(33-2N_f)$ denoting the one-loop anomalous dimension of the running mass, and $\Lambda _{\rm QCD}=343(12)\ \text{MeV}$ is taken from the lattice QCD simulation result~\cite{FlavourLatticeAveragingGroup:2019iem}.
We then take Eq.~(\ref{RGI}) to determine the condensate in the chiral limit at the lattice QCD renormalization scale ${{\zeta }_{lat}}=2\ \text{GeV}$,
which gives ${{\left\langle \bar{q}q \right\rangle }_{\chi }}({{\zeta }_{lat}})=-{{(273.1\ \text{MeV})}^{3}}$, agrees excellently with the lattice QCD results $-{{(272(5)\ \text{MeV})}^{3}}$~\cite{Bazavov:2010yq,Boyle:2015exm}.
	
In case of beyond chiral limit, it has been well known that the quark condensate involves a quadratic divergence. Such a divergence in the condensate can be regularized by considering the variation of the current quark mass:
\begin{equation}\label{sbt_Cd}
{{\left\langle \bar{q}q \right\rangle }_{m,\zeta }}= - \Big{(} 1 - {{m}^{\zeta }}\frac{\partial }{\partial {{m}^{\zeta }}} \Big{)} \text{Tr}[S].
\end{equation}
On the other hand, the chiral condensate is related to the physical observables, e.g. the pion mass $m_\pi$, via the Gell-Mann-Oakes-Renner (GOR) relation:
\begin{equation}    \label{GOR}
f_{\pi} ^{2} m_{\pi} ^{2}  = - 2{m_{l,\zeta}^{}}\, \left\langle {\bar qq} \right\rangle_{m,\zeta}^{} \, .
\end{equation}
For the pion decay constant $f_{\pi}$, we apply the Pagels-Stokar formula~\cite{Pagels:1979hd,Roberts:1994dr}:
\begin{align}
f_{\pi} ^{2} = & \, \frac{3}{{4{\pi ^2}}}\int_0^\Lambda  d{p^2}\frac{{{p^2}Z\left( {{p^2}} \right){M}\left( {{p^2}} \right)}}{{{{\left[ {{p^2} + {M^2}\left( {{p^2}} \right)} \right]}^2}}} \notag\\
	& \, \times \left[ {M\left( {{p^2}} \right) - \frac{{{p^2}}}{2}\frac{{dM\left( {{p^2}} \right)}}{{d{p^2}}}} \right],
\end{align}
with:
\begin{align}
Z(p^2) &= 1/A(p^2), \\
M(p^2) &= B(p^2)/A(p^2). \label{eq:quarkmass}
\end{align}
We get $f_\pi = 91.5\,\rm{MeV}$ for the minimal \textit{Ans\"atze} in light quark case,
which is comparable with the experimental data ${{f}_{\pi ,phy}}=92.1\ \text{MeV}$~{\cite{ParticleDataGroup:2018ovx}}.
The current quark mass is then constrained with Eq.~(\ref{GOR}) by taking the experimental value of the pion mass ${{m}_{\pi ,phy}}=138\ \text{MeV}$ and the calculated pion decay constant ${{f}_{\pi ,phy}}=92.1\ \text{MeV}$.
The current mass of light quark is then ${{m}_{l,\zeta}^{}}=2.7\ \text{MeV}$, along with the s quark mass  ${{m}_{s}}/{{m}_{l}}=27$.
	
\section{Numerical result}
\label{sec:4}

With our presently established framework, we calculate the effective potential $\Gamma_{\Sigma }$ of QCD system in both vacuum and finite temperature and density cases, and investigate the phase coexistence region of QCD.
%

	
\subsection{In the vacuum}

\begin{figure}[htbp]
\centering
\includegraphics[width=7.5cm]{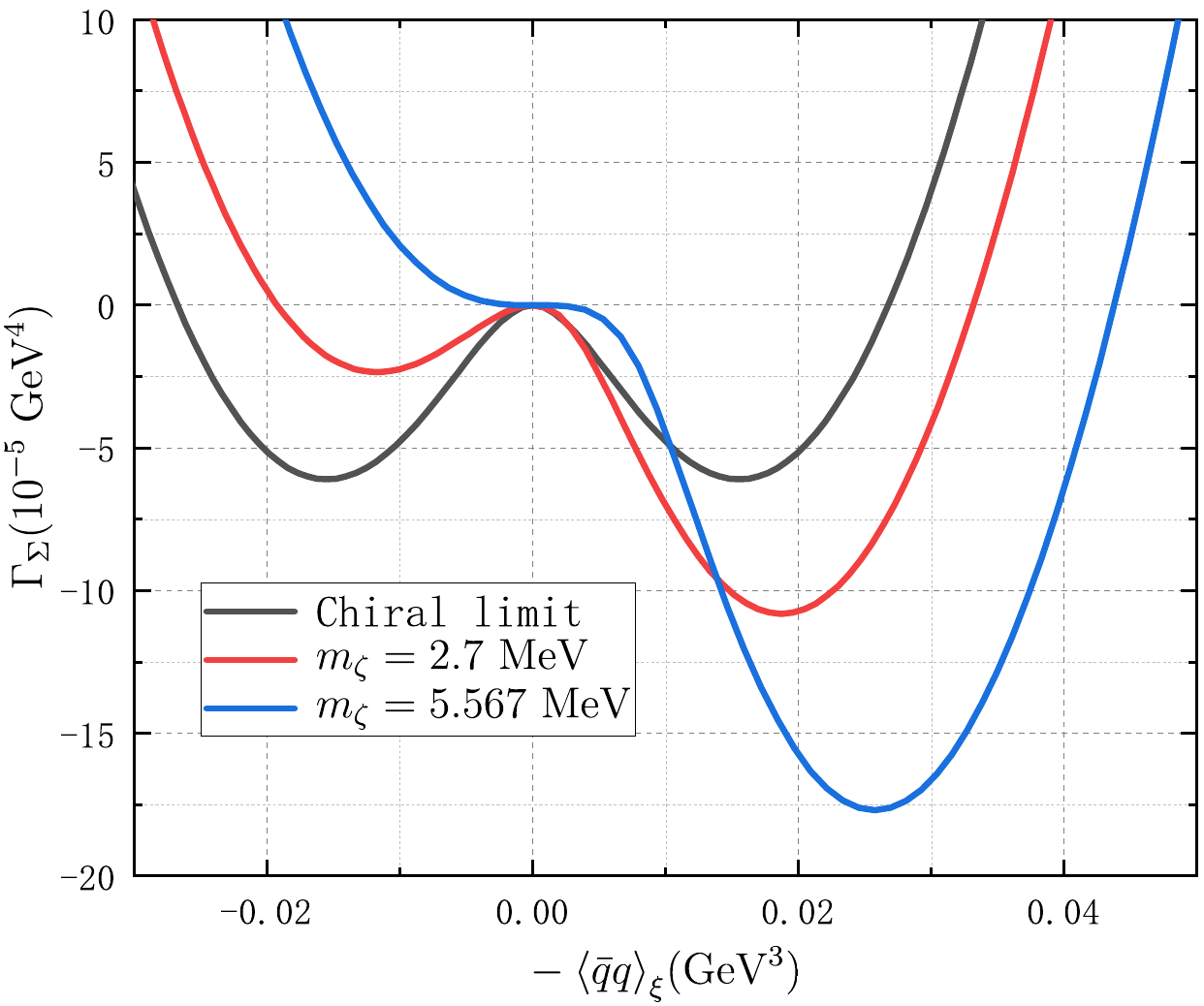}
\vspace*{-3mm}
\caption{The obtained effective potential $\Gamma_{\Sigma }$ as a function of the RGI quark condensate, in the homotopy transformation between the Wigner and the Nambu$\pm$ solutions,
in cases of several current quark masses $m_{\zeta}$. Here we have defined the effective potential to vanish at $\langle\bar q q\rangle_{\zeta}^{} = 0$, which is the condensate of the Wigner solution. }
\label{fig:eff_mass}
\end{figure}

It has been well known that, in the vacuum, the gap equation has the Nambu$\pm$ and the Wigner solutions, which are the representation of the chiral symmetry dynamical breaking phase, dynamical chiral symmetry phase, respectively, in case of small current quark mass (beyond the chiral limit).
To calculate the effective action from Eq.~(\ref{AF_first_deri}), it is natural to choose the optimized direction, \textit{i.e.}, the homotopy between the Nambu$\pm$ and Wigner solutions.
Following Eq.~(\ref{eq:homoS}), one can define:
\begin{equation}\label{homo}
S^{-1}_{\xi_{\pm}} =(1 -\xi_{\pm}^{} ) S_{W}^{-1} + \xi_{\pm} S_{N\pm}^{-1} \, ,
\end{equation}
where $S_{N\pm}$ and $S_{W}$ are the propagators of the Nambu$\pm$, the Wigner solutions, respectively.
$\xi_{\pm}^{}$ serves as the homotopy parameter between the Nambu$\pm$ and Wigner solutions.
In other words, a homotopy trajectory $\xi_{+}^{}$ is chosen to connect the Wigner solution and the Nambu$+$ solution, and another homotopy trajectory $\xi_{-}^{}$ connects the Wigner solution and the Nambu$-$ solution.
Then, we can find the difference of the effective potential between any two homotopy functions specified by Eq.~(\ref{homo}).
Meanwhile, the ${\xi}$ dependent chiral condensate $-\langle\bar q q\rangle_{\zeta}^{}$ can be calculated from the corresponding propagator $S_{\xi}$ by defining the Wigner chiral condensate to vanish,
for the case of $\xi_{+}$ and also for $\xi_{-}$.
%
%
In the following we will simply denote $\langle\bar q q\rangle _{\xi}$ for the two cases.
We then obtain the effective potential as a function of the order parameter, \textit{i.e.}, the RGI condensate $\langle\bar q q\rangle_{\zeta}^{}$.
The obtained result is depicted in Fig.~\ref{fig:eff_mass}.

\begin{figure}[htbp]
\centering
\includegraphics[width=7.5cm]{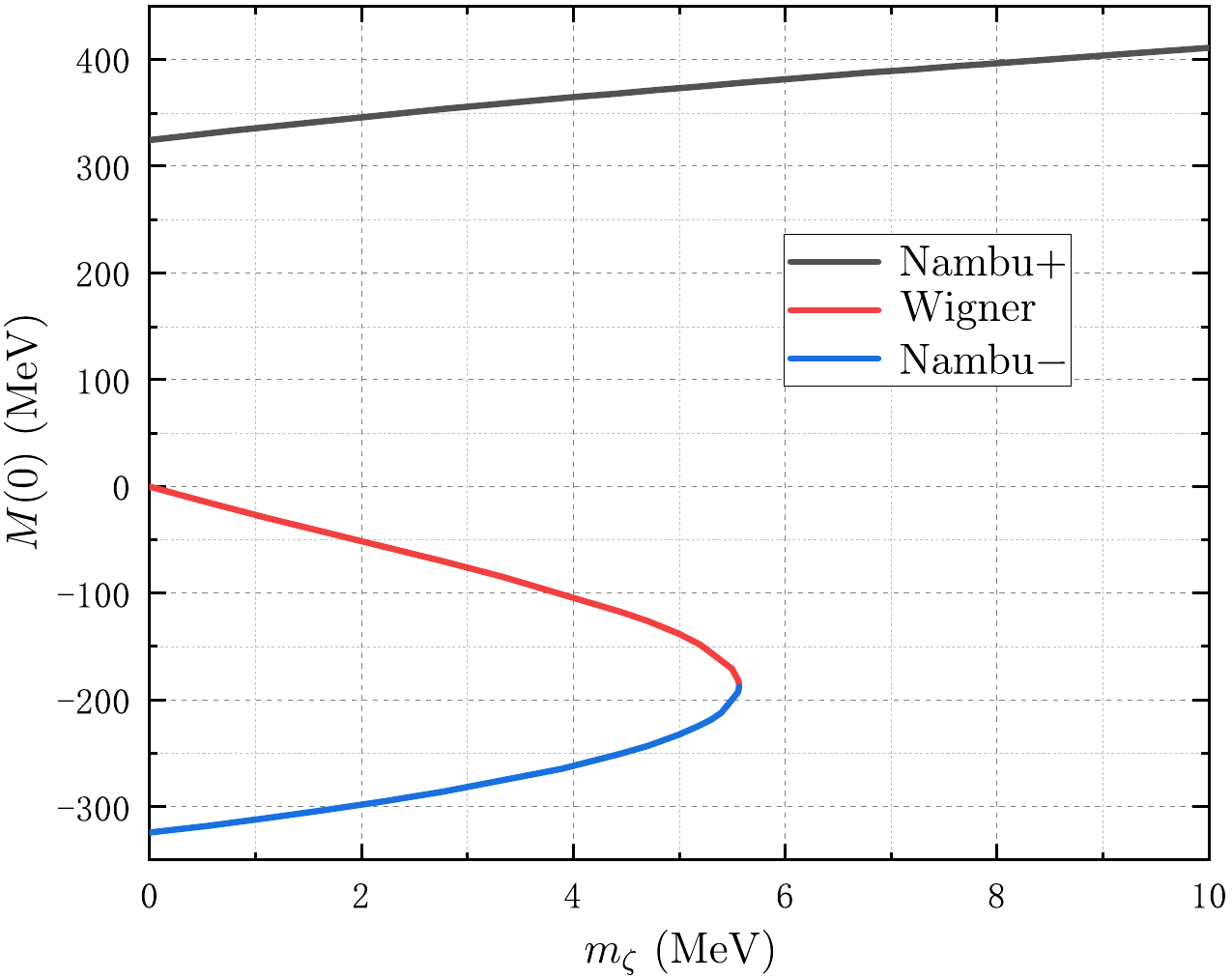}
\vspace*{-3mm}
\caption{The obtained structure of the multiple solutions of the quark gap equation with the minimal vertex \textit{Ans\"atze}, which is demonstrated by the current quark mass dependence of the $M(0)$ defined by Eq.~(\ref{eq:quarkmass}).
The black, red and blue solid curves corresponds to the Nambu$+$, Wigner and Nambu$-$ solutions, respectively.}
\label{fig:M0}
\end{figure}

From Fig.~\ref{fig:eff_mass}, we can observe evidently that the order-parameter dependence of the $\Gamma_{\Sigma }$ at different current quark masses
and find that the multiple solutions vanish at a critical mass ${{m}_{crit}}=5.567\ \text{MeV}$.
in case of $m < m_{crit}$, the Wigner solution exhibits as the maximum in the effective potential, while the Nambu$+$ solution corresponds to a minimum, indicating that the physical solution is the Nambu$+$ solution.
In the chiral limit, the QCD vacuum exhibits a degeneracy for the Nambu$+$ and the Nambu$-$ solutions.
Beyond the chiral limit with a mass smaller than the critical one, physical solutions appear in regions where the condensate is negative, known as Nambu$+$ solutions~\cite{Roberts:1994dr,Roberts:2000aa,Wang:2012me}.
At the critical mass ${{m}_{crit}}=5.567\ \text{MeV}$, the effective potential shows an inflection point at the Wigner solution, which suggests the merging between the Wigner and the Nambu$-$ solutions.
This critical mass coincides with the critical mass obtained from the structure of the multiple solutions in the quark gap equation. The numerical result of the solution structure is shown in Fig.~\ref{fig:M0},
which is obtained by solving the quark DSE (gap equation) with the standard iteration process.
	
\subsection{At Finite temperature and density}

We now study the phase diagram of QCD at finite temperature and baryon density (chemical potential).
It has been estimated that the chiral phase transition at a low baryon density is a crossover and becomes a first-order transition in case of a larger chemical potential, with a critical end point (CEP) linking them. To study the (pseudo-)critical temperature of the chiral phase transition ${{T}_{c}}({{\mu }_{B}})$, we consider the reduced condensate~\cite{Cheng:2007jq,Fischer:2014ata}:
\begin{equation}
{{\left\langle \bar{q}q \right\rangle }_{l,s}}={{\left\langle \bar{q}q \right\rangle }_{l}}-\frac{{{m}_{l}}}{{{m}_{s}}}{{\left\langle \bar{q}q \right\rangle }_{s}},
\end{equation}
and define the $T_{c}$ as the one corresponding to the peak of the thermal susceptibility:
\begin{equation}
\chi_{T}^{} = \frac{\partial {\left\langle \bar{q}q \right\rangle }_{l,s} }{\partial T} \, .
\end{equation}

	
\begin{figure}[htbp]
\centering
\includegraphics[width=7.5cm]{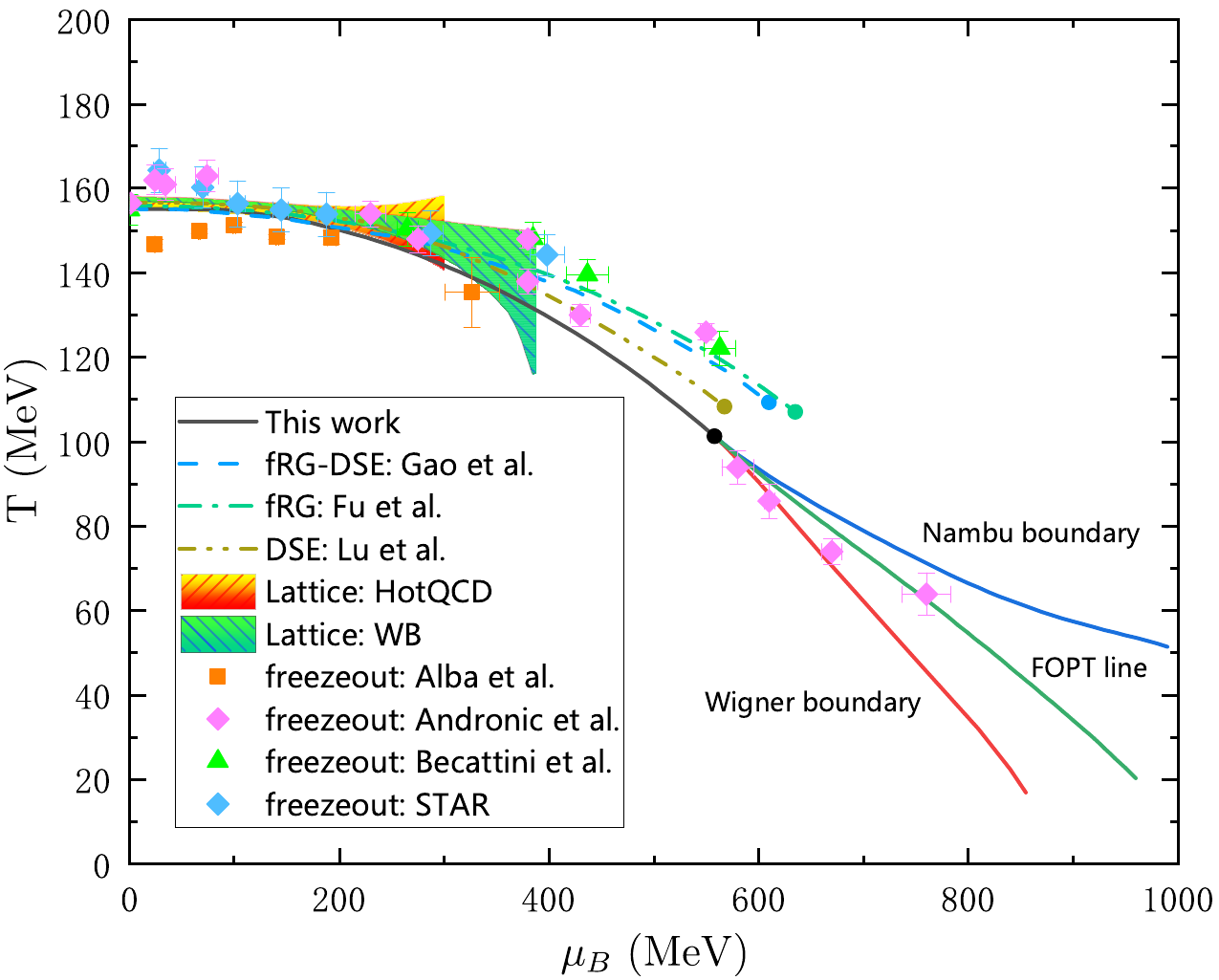}
\vspace*{-3mm}
\caption{Phase diagram for the 2+1-flavour QCD. This work: black solid - chiral phase transition line (crossover); red solid - phase boundary where the Wigner solution disappears; blue solid - phase boundary where the Nambu solution disappears; green solid - the first-order phase transition line, where the effective potentials of the Nambu and Wigner phases are equal to each other. We also display the chiral phase transition line from other functional QCD studies~\cite{Fu:2019hdw,Gao:2020qsj,Lu:2023mkn} and from the lattice QCD extrapolation~\cite{Borsanyi:2020fev,HotQCD:2018pds}, together with the extracted freeze-out data from different groups~\cite{Alba:2014eba,Becattini:2016xct,STAR:2017sal,Andronic:2017pug}. }
\label{fig:phase}
\end{figure}

With the above criterion, we obtain that the pseudo-critical temperature of the chiral phase transition at $\mu_{B}^{} =0$ is $T_{c} = 155.2\, \rm{MeV}$, which is apparently consistent with the lattice QCD results $T_c = 156.5(\pm1.5)\ \rm{MeV}\,$~\cite{HotQCD:2019xnw}.
For the chiral phase transition temperature with an increasing of the baryon chemical potential, our results can be well described by the following equation:
\begin{equation}
\frac{{{T}_{c}}({{\mu }_{B}})}{{{T}_{c}}}=1-\kappa {{\left( \frac{{{\mu }_{B}}}{{{T}_{c}}} \right)}^{2}}-\lambda {{\left( \frac{{{\mu }_{B}}}{{{T}_{c}}} \right)}^{4}}+\cdots, \label{eq:PTline-4}
\end{equation}
where the $\kappa$ stands for the curvature of the phase transition line.
%
%
%
The parameters of the phase transition line in \Eq{eq:PTline-4} are found to be:
\begin{equation}
\kappa =0.0184 \, , \quad \lambda = 0.00128 \, .
\end{equation}
Our result for the curvature $\kappa$ is slightly larger than recent lattice QCD results~\cite{Borsanyi:2020fev,Guenther:2020jwe}, but still matches with that given by earlier lattice QCD, $\kappa =0.018(4)$~\cite{Cea:2014xva} and those from the functional QCD approaches~\cite{Fu:2019hdw,Gunkel:2021oya,Gao:2020fbl}. $\lambda$ can also be comparable with the lattice QCD, $\lambda=0.000(4)$ \cite{HotQCD:2018pds}.
Using the thermal susceptibility, we also get the critical end point (CEP) to be located at:
\begin{equation}
({T^{E}, \, {\mu }^E_{B}})\ = (101.3 \, , \,  558.0)\, \rm{MeV} \, .
\end{equation}
which is also comparable with the current estimation of the CEP location \cite{Fu:2019hdw,Gunkel:2021oya,Gao:2020fbl,Lu:2023mkn,Hippert:2023bel}.
To summarize, we show our calculated chiral phase transition line in Fig.~\ref{fig:phase},
which is also compared with the results from functional QCD studies~\cite{Fu:2019hdw,Gao:2020qsj,Lu:2023mkn}, the lattice QCD extrapolation~\cite{Borsanyi:2020fev,HotQCD:2018pds},
and also the extracted freeze-out data from several groups~\cite{Alba:2014eba,Becattini:2016xct,STAR:2017sal,Andronic:2017pug}.
We note that the difference between our present result and the one in Ref.~\cite{Lu:2023mkn} is mainly due to the different treatment on the quark loop contribution to gluon propagator,
here it is simply computed with the hard thermal-loop approximation result in Eq.~(\ref{eq:htl-qrk}).

\begin{figure}[htbp]
\centering
\includegraphics[width=7.5cm]{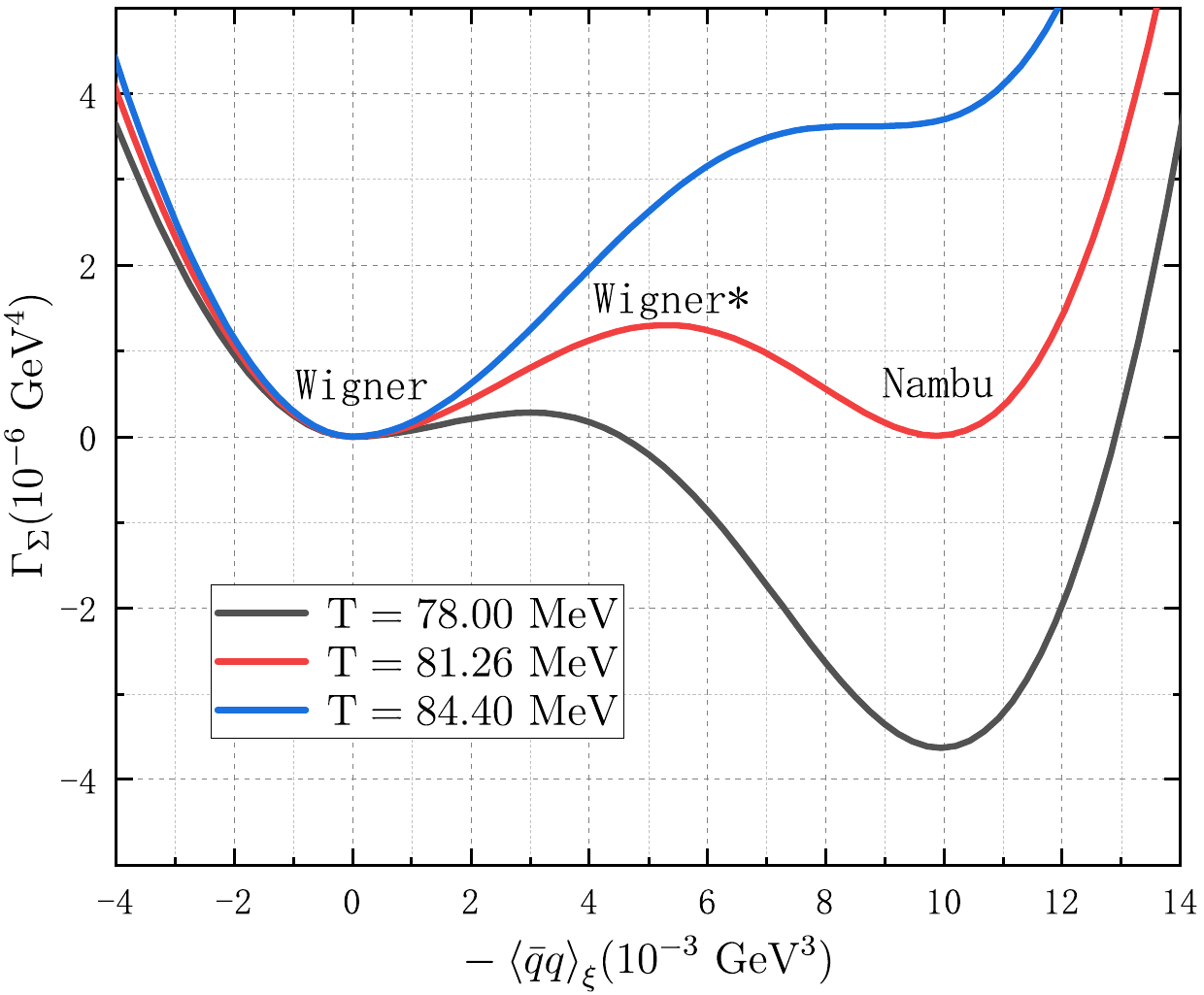}
\vspace*{-3mm}
\caption{The obtained effective potential at a given baryon chemical potential $\mu_{B}=660\ \rm{MeV}$ and several values of temperature, which demonstrates the evolution of the physical phase between the Nambu$+$ and the Wigner solutions.
We have again defined the effective potential to vanish at $\langle\bar q q\rangle_{\zeta}^{} = 0$.}
\label{fig:FOPT}
\end{figure}

With the framework described in the last section, we can calculate the effective potential as a function of the RGI quark condensate $\langle\bar q q\rangle_{\zeta}^{} $ of the QCD system.
The obtained results in the cases of a larger given chemical potential ($660\, \rm{MeV}$) and some values of lower temperature are displayed in Fig.~\ref{fig:FOPT}.
It is then possible to determine the first-order phase transition line with conventional effective potential criterion,
which reads that, on the phase transition line (the solid green one in Fig.~\ref{fig:phase}) the minimum of the effective potential in Eq.~(\ref{AF_first_deri}) of the two phases equate to each other, \textit{i.e.}, with two degenerate minima.
Our obtained result is shown as the solid red line in Fig.~\ref{fig:FOPT} (Corresponding to $\mu_{B}^{} = 660\, \rm{MeV}$, $T_{c} = 81.26\, \rm{MeV}$).
It is also evident that there exist regions for both the Nambu and the Wigner solutions to coexist in the first-order phase transition region.
%
%
It is known that, below the Wigner boundary (the solid red line in Fig.~\ref{fig:phase}), there only exists the Nambu phase, while above the Nambu boundary (the solid blue line in Fig.~\ref{fig:phase}), only the Wigner phase is left,
in the between both the Nambu and the Wigner solutions coexist.
The Fig.~\ref{fig:FOPT} manifest this feature in more detail.
When the temperature is below the critical temperature, the Nambu phase is the stable physical phase and the Wigner phase is the meta-stable phase (e.g., the case at $T = 78.00\,\textrm{MeV}$),
otherwise, the Wigner phase is the stable physical one (e.g., the case at $T = 84.40\,\textrm{MeV}$).
	
\begin{figure}[htbp]
\centering
\includegraphics[width=8cm]{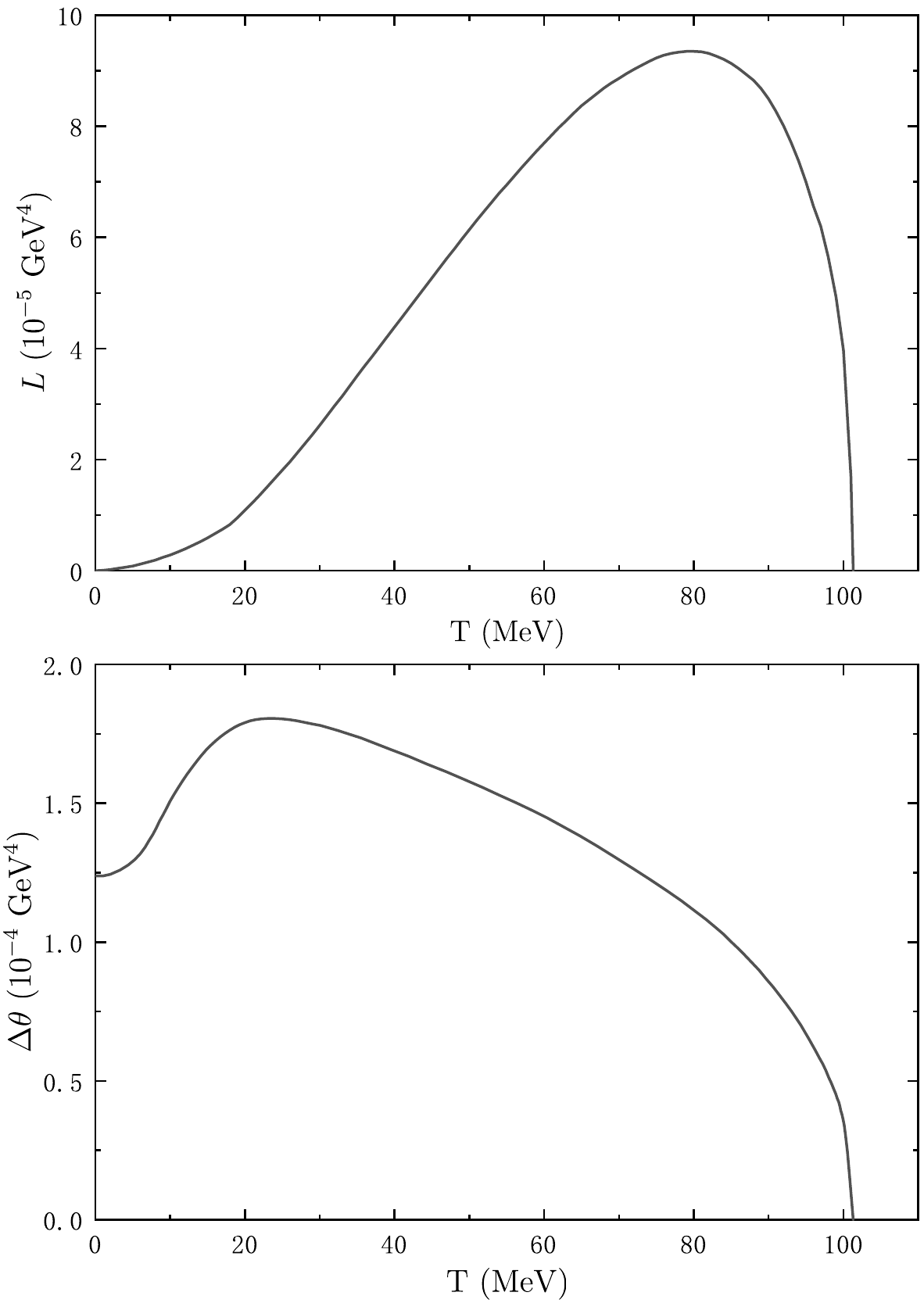}
\vspace*{-3mm}
\caption{The obtained temperature dependence of the latent heat density and the difference of trace anomaly, {evaluated} on the first-order phase transition line in the QCD phase diagram. Upper panel: the latent heat density $L$. Lower panel: the difference of trace anomaly between two phases $\Delta\theta$.}
\label{fig:lheat_trano}
\end{figure}

Moreover, with our developed framework, we can obtain directly the temperature dependence of the pressure difference between the Nambu and the Wigner solutions, which allows us to calculate the latent heat density $L$ and  the difference of trace anomaly between two phases $\Delta\theta$~\cite{Gao:2023djs,Hindmarsh:2015qta,Kamionkowski:1993fg} which are the typical characteristics of the first-order phase transition line. 
They can be written in terms of the pressure as:
\begin{align}
L& =T(s_{W}^{} - s_{N}^{} ) = T\left[ \left(\frac{\partial P_{W} }{\partial T}\right)_{\mu_{B}^{}} -\left(\frac{\partial P_{N} }{\partial T}\right)_{\mu_{B}^{}} \right] \, , \label{latent_heat} \\
\Delta\theta & =\frac{1}{4}  L + \frac{1}{4} \mu_{q} \left(n_{q}^{W} - n_{q}^{N} \right)-(P_{W} - P_{N} )  \, , \label{Delta I}
\end{align}
with $P$ the pressure where $P=-\Gamma[\Sigma]$, $s$ the entropy density and ${{n}_{q}}=\frac{\partial P}{\partial \mu_q}$ the quark number density.
We then obtain the temperature dependence of the latent heat density $L$ and the difference of the trace anomaly  $\Delta\theta$  of the first-order phase transition,
as shown in Fig.~\ref{fig:lheat_trano}.
It is apparent that, at $T=0$, the latent heat density is zero since the temperature in Eq.~(\ref{latent_heat}) equals zero.
And at $T=101.3\,\rm{MeV}$, \textit{i.e.}, the temperature of the CEP, the latent heat density vanishes, because the derivative of the pressure difference becomes zero at the CEP, indicating the first-order phase transition terminates.
Meanwhile a maximum appears at $T\approx 80\,\rm{MeV}$, at which the latent heat density $L=9.36\times 10^{-5}\,\rm{GeV}$ is approximately equal to the pressure difference of the QCD vacuum between the Wigner and the Nambu solution $\Delta\Gamma_\Sigma=10.81\times 10^{-5}\,\rm{GeV}$.
On the other hand, the difference of the trace anomaly $\Delta\theta$ has a peak at $T\approx 24\,\rm{MeV}$.
Below this temperature, the difference of the number densities, \textit{i.e.} the second term on the right hand side of Eq.~(\ref{Delta I}), contributes most to $\Delta\theta$. 
Since the quark mass in the Wigner phase grows as temperature decreases, its number density $n_q^W$ gets suppressed and hence lowers the trace anomaly at very low temperature.
At $T=101.3\,\rm{MeV}$, $\Delta\theta$ vanishes at the temperature of the CEP, since the first-order phase transition ends.
%
Finally, the latent heat density and the difference of trace anomaly on the first-order phase transition line can be applied to study the dynamical evolution in heavy-ion collisions and also the universe evolution, which is under progress.

\section{summary}
\label{sec:5}

In this paper, we develop a homotopy method to compute the effective potential of QCD, which allows us to compute the potential from the gap equation under any kind of  truncation scheme.
Here we apply an optimized minimal structure for the truncation of the quark-gluon interaction vertex, which provides consistent results compared to the lattice and the functional QCD approaches.
We also construct a new effective action, which generalizes the CJT connected generating functional by the generalized Legendre transformation.
Under the transformation, the effective action is naturally converted to a functional of the self-energy, instead of the propagator, thereby avoiding the saddle point issue in the CJT action.
Besides, the functional derivative of the newly defined effective potential is a generalized equation of motion, which is an inverse form of the equation of motion of the CJT action,
 thus allowing one to obtain the potential with the lower bound constraints.

Building upon the homotopy method and the new effective potential, we calculate the current quark mass dependence of the effective potential in the QCD vacuum,
and observe a merging behavior of the Nambu$-$ and Wigner solutions at the critical quark mass where the effective potential exhibits as an inflection point.
The obtained critical quark mass is consistent with the result from solving the gap equation with standard iteration process.
Additionally, for the current quark mass above the critical one, the Nambu$-$ and Wigner solutions vanish, leaving only the Nambu$+$ solution, which includes both the DCSB and the explicit chiral symmetry breaking effect.

Moreover, we study the first-order phase transition of QCD system at finite temperature and large baryon chemical potential, and obtain the phase diagram with the first-order phase transition line, where the minimum of the effective potential of the two phases equate to each other. 
In the region of the first-order phase transition, both the Wigner and the Nambu$+$ phases appear as the minima in the effective potential, with a maximum locating in the between,
indicating that our presently obtained effective potential captures accurately the behavior of the first-order phase transition.
On the first-order phase transition line,
we also obtained the temperature dependence of the latent heat density and the difference of trace anomaly.
%
%
This can be further incorporated in the determination of the dynamical evolution process of the phase transition, for instance, in the evolution of the fireball produced relativistic heavy-ion collisions and in the universe evolution.
The work is under progress.



\vspace{\baselineskip}
\vspace{\baselineskip}
\begin{acknowledgments}
This work is supported by the National  Science Foundation of China under Grants  No.12175007,  No.12247107 and No.12305134.
\end{acknowledgments}

	\bibliography{eff_potent}
\end{document}